\documentstyle[12pt]{article}

\advance \topmargin by -\headheight
\advance \topmargin by -\headsep

\evensidemargin \oddsidemargin
\newcommand{\ggg}{{g_{11}}}
\newcommand{\gggg}{{ g_{01}}}
\newcommand{\ggggg}{{g_{00}}}
\newcommand{\aaa}{{\gamma_{11}}}
\newcommand{\aaaa}{{\gamma_{01}}}
\newcommand{\aaaaa}{{\gamma_{00}}}
\newcommand{\sss}{{\sqrt{-g}}}

\begin{document}
\font\ninerm = cmr9

\def\footnoterule{\kern-3pt \hrule width \hsize \kern2.5pt}

\pagestyle{empty}
\begin{center}
{\large\bf Area-preserving Structure and Anomalies in 1+1-dimensional
Quantum Gravity}%
\footnote{\ninerm This work is supported
in part by funds provided by the U.S. Department of Energy (D.O.E.)
under cooperative agreement \#DE-FC02-94ER40818, as well as in part
by the National Science Foundation under contracts \#INT-910559 and
\#INT-910653, and by Istituto Nazionale di
Fisica Nucleare (INFN, Frascati, Italy). }

\vskip 1cm
G. Amelino-Camelia, D. Bak, and D. Seminara
\vskip 0.5cm
{\it Center for Theoretical Physics\\
Laboratory for  Nuclear Science and Department of Physics\\
Building 6, Massachusetts Institute of Technology\\
Cambridge, Massachusetts 02139, U.S.A.}

\end{center}

\vspace{1.2cm}
\begin{center}
{\bf ABSTRACT}
\end{center}

{\leftskip=0.6in \rightskip=0.6in

We investigate the gauge-independent Hamiltonian
formulation and the anomalous
Ward identities of a matter-induced
1+1-dimensional gravity theory
invariant under Weyl transformations and area-preserving diffeomorphisms,
and compare the results to the ones for the conventional
diffeomorphism-invariant theory.
We find that, in spite of several
technical differences encountered in the analysis,
the two theories are essentially equivalent.
}

\vskip 3cm
\centerline{Submitted to: Physics Letters B}

\vfill

\hbox to \hsize{MIT-CTP-2429  \hfil May 1995}

\newpage
\baselineskip 12pt plus .5pt minus .5pt
\pagenumbering{arabic}
\pagestyle{plain}

\section{Introduction}
It is well known\cite{pol81} that in
the quantization of the 1+1-dimensional matter-gravity field theory
with action
\begin{eqnarray}
{\cal I}(X,g) = {1 \over 2} \int d^2\xi ~ \sqrt{-g} ~
g^{\mu \nu} ~ \partial_\mu X_A \, \partial_\nu X^A ~,
\label{sxg}
\end{eqnarray}
where $g_{\mu \nu}$ is a metric tensor
with signature (1,-1), $A \! = \! 1,2,...,d$,
and $X$ is a d-component massless scalar field,
one encounters anomalies that break part of
the symmetry of the classical theory, which, as seen from ${\cal I}(X,g)$,
has Weyl and diffeomorphism invariance.

In the conventional quantization
approach[1-8]
diffeomorphism invariance is preserved,
while  renouncing Weyl invariance.
In the path integral formulation this can be accomplished by choosing
the following measures for the functional
integrations\cite{dis88,dav88}
\begin{eqnarray}
\int {\cal D}\delta X ~ {\rm exp}\left(i \int d^2\xi ~ \sqrt{-g} ~
\delta X_A \,  \delta X^A \right) ~ = ~ 1
{}~, \label{measpone}\\
\int {\cal D}\delta g ~ {\rm exp} \left(i \int d^2\xi ~ \sqrt{-g} ~
g^{\mu \sigma} \, g^{\nu \rho} ~
\delta g_{\mu \nu} \, \delta g_{\sigma \rho}~ \right) = ~ 1
{}~, \label{measptwo}
\end{eqnarray}
which are diffeomorphism invariant, but
are not Weyl-invariant.

Integrating out the
matter degrees of freedom using the
measure (\ref{measpone}) one obtains
an effective pure gravity theory
with action
\begin{eqnarray}
\Gamma^D(g) \! = \! - \Lambda^D \! \int \! d^2\xi
\sqrt{-g(\xi)} + {d \over 96 \pi} \!
\int \! d^2\xi_1 \, d^2\xi_2 \, \sqrt{-g(\xi_1)} \,
R(g(\xi_1)) ~ \Box^{-1}(\xi_1,\xi_2) ~ \sqrt{-g(\xi_2)} \, R(g(\xi_2))
\, , \label{sgp}
\end{eqnarray}
where $\Box^{-1}$ is the inverse of the Laplace-Beltrami operator,
and the index $D$ stands for ``diffeomorphism invariant approach".
The cosmological term $\Lambda^D \!\int \! d^2\xi \, \sqrt{-g}$
is induced by renormalization\cite{pol81}.

As it should be expected, $\Gamma^D(g)$ is
diffeomorphism-invariant, but
it is not Weyl-invariant; in fact, the energy-momentum tensor
\begin{eqnarray}
T^D_{\mu \nu}  \equiv {2 \over \sqrt{-g}}
{\delta \Gamma^D(g) \over \delta g^{\mu \nu}} ~.
\label{tp}
\end{eqnarray}
is covariantly conserved, but possesses non-vanishing trace
\begin{eqnarray}
\nabla_\mu (g^{\mu \nu} T^D_{\nu \alpha}) = 0  ~,
{}~~~~~g^{\mu \nu} T^D_{\mu \nu} = 2 \Lambda^D + {d \over 24 \pi} R(g) ~.
\label{anop}
\end{eqnarray}

Recently, an alternative approach to the quantization of the classical
theory of action (\ref{sxg})
has been considered\cite{kar94,jac95},
in which part of the diffemorphism invariance is
sacrificed
in order to obtain a
Weyl-invariant theory.
This alternative approach is motivated by the observation
that Eq.(\ref{sxg})
depends on the metric only through the Weyl-invariant combination
\begin{eqnarray}
\gamma^{\mu \nu} \equiv \sqrt{- g} \, g^{\mu \nu}
{}~. \label{gamma}
\end{eqnarray}
In the
conventional
diffeomorphism-invariant
quantization approach
one is forced to introduce,
through the path integral measures,
a field
describing the determinant of the metric $g$,
even though (\ref{gamma}) is independent of this determinant.
Instead, the Weyl-invariant approach does
not lead to the introduction of this
additional field and might therefore be a more natural\cite{jac95}
candidate as a quantized version of the original classical theory.

In the path integral formulation of this alternative Weyl-invariant
quantization one can choose the following measures
for the functional integrations
\begin{eqnarray}
\int {\cal D}\delta X ~ {\rm exp}\left(i \int d^2\xi ~ \,
\delta X_A \,  \delta X^A \right) ~ = ~ 1
{}~, \label{measjone}\\
\int {\cal D}\delta \gamma ~ {\rm exp}\left(i \int d^2\xi ~
\gamma^{\mu \sigma} \gamma^{\nu \rho} \,
\delta \gamma_{\mu \nu} \delta \gamma_{\sigma \rho}~ \right) = ~ 1
{}~, \label{measjtwo}
\end{eqnarray}
which depend on $g$ only through $\gamma$,
and can be obtained from
(\ref{measpone}) and (\ref{measptwo}) with the substitution
$g \rightarrow \gamma$ (also observing that $\det \gamma = -1$).

The measures (\ref{measjone}) and (\ref{measjtwo})
are evidently Weyl invariant, and are also invariant
under the subgroup of the diffeomorphism group
given by coordinate redefinitions with unit Jacobian,
{\it i.e.} infinitesimally
$\xi^\mu \rightarrow \xi^\mu + f^\mu$ with $\partial_\mu f^\mu = 0$,
which are called area-preserving
diffeomorphisms
because they
preserve local area $\sqrt{-g} d^2\xi$ on spaces where $\sqrt{-g}$
is constant.
However, (\ref{measjone}) and (\ref{measjtwo}) are not
invariant under diffeomorphisms with
non-unit Jacobian, and therefore, as already pointed out in \cite{jac95},
this Weyl-invariant approach has
as many symmetries as the diffeomorphism-invariant
approach\footnote{Note that, 1+1-dimensional
area-preserving diffeomorphisms are parametrized by
one arbitrary function (in 1+1 dimensions $\partial_\mu f^\mu = 0$
is locally solved by $f^\mu = \epsilon^{\mu \nu} \partial_\nu \phi$),
and another arbitrary function is needed to parametrize
Weyl transformations,
whereas 1+1-dimensional
diffeomorphisms are parametrized by
two arbitrary functions.}.

The effective quantum gravity theory
that follows from the action (\ref{sxg}) upon integration of $X$
using the measure (\ref{measjone}) is described by the
action\cite{kar94,jac95}
\begin{eqnarray}
\Gamma^W(\gamma) \! = \! - \Lambda^W \! \left(\int \! d^2\xi \right)
+ {d \over 96 \pi}
\int d^2\xi_1 \, d^2\xi_2 ~  \, R(\gamma(\xi_1)) \,
\Box^{-1}(\xi_1,\xi_2) \,  R(\gamma(\xi_2)) ~,
\label{sgj}
\end{eqnarray}
which is
Weyl invariant, but
is not diffeomorphism invariant; in fact
\begin{eqnarray}
T^W_{\mu \nu} ~ \equiv ~ {2 \over \sqrt{-g}}
{\delta \Gamma^W(\gamma) \over \delta g^{\mu \nu}}
{}~ = ~ 2 {\delta \Gamma^W(\gamma) \over \delta \gamma^{\mu \nu}} -
\gamma_{\mu \nu} \gamma^{\alpha \beta}
{\delta \Gamma^W(\gamma) \over \delta \gamma^{\alpha \beta}}
\label{tj}
\end{eqnarray}
satisfies the following anomaly relations
\begin{eqnarray}
\nabla_\mu (g^{\mu \nu} T^W_{\nu \alpha}) = - {d \over 48 \pi}
{1 \over \sqrt{-g}} \partial_\alpha R(\gamma) ~,
{}~~~~~g^{\mu \nu} T^W_{\mu \nu} = 0 ~.
\label{anoj}
\end{eqnarray}
The term $\Lambda^W \!\int \! d^2\xi$
is induced by renormalization, and,
even though it is $\gamma$-independent (and obviously
does not contribute to the anomaly relations),
can have a non-trivial role in the theory since it
gives
different weights to surfaces with different $\int \! d^2\xi$
in the evaluation of the partition function.

\noindent
Also notice that the anomaly relations (\ref{anoj}) can be put
in the following $\sqrt{-g}$-independent form
\begin{eqnarray}
\hat\nabla_\mu (\gamma^{\mu \nu} T^W_{\nu \alpha}) =
- {d \over 48 \pi} \partial_\alpha R(\gamma) ~,
{}~~~~~\gamma^{\mu \nu} T^W_{\mu \nu} = 0 ~,
\label{anojgamma}
\end{eqnarray}
where $\hat\nabla$ is the
covariant derivative computed with the metric $\gamma_{\mu\nu}$, and
the invariance of $\Gamma^W(\gamma)$ under area-preserving
diffeomorphism is encoded in the fact that\cite{jac95}
\begin{eqnarray}
\hat\nabla_\mu \hat\nabla_\nu (\gamma^{\beta \nu} \epsilon^{\mu \alpha}
T^W_{\alpha \beta}) = 0 ~.
\label{anojsdiffeo}
\end{eqnarray}

In this Letter, we
investigate this alternative Weyl-invariant approach both in
the gauge-independent Hamiltonian
formulation\cite{abd91} and using anomalous
Ward identities\cite{pol87}, and compare the results
to the ones for the conventional diffeomorphism invariant approach.

\section{Classical Hamiltonian Formulation}
In this section we discuss\footnote{Note that the Hamiltonian
formulation of the conventional diffeomorphism-invariant approach
was already discussed in Ref.\cite{abd91}.}
the classical Hamiltonian formulation of both approaches
reviewed in the Introduction, and compare their respective
simplectic structure and constraints.
This type of Hamiltonian analysis should be affected very strongly
by the different symmetry properties of the two approaches.
In the diffeomorphism-invariant approach
the symmetries impose that the Hamiltonian (which generates
diffeomorphisms along the time direction) weakly vanish on the
surface defined by the two diffeomorphism constraints.
On the other hand,
in the Weyl-invariant approach the fact that the Weyl symmetry
is already enforced by working with $\gamma$ rather than $g$
suggests that the constraint surface be
determined only by the constraint enforcing invariance under
area-preserving diffeomorphisms.
This should lead to a closed
constraint algebra because the area-preserving diffeomorphisms
form a group.
Moreover,
since diffeomorphisms in the time direction
are not area-preserving, there might be room for
a non-vanishing Hamiltonian.

We work with the localized versions\cite{abd91,abdbook}
of the nonlocal effective actions in (\ref{sgp})
and (\ref{sgj}), which are respectively given
by\footnote{In analogy with the argument used
in the proof\cite{abdbook}
of the equivalence of the diffeomorphism-invariant actions
$S^D$ and $\Gamma^D$, we have verified the equivalence
of the Weyl-invariant actions
$S^W$ and $\Gamma^W$ by integrating out
the scalar field $\varphi$, and
evaluating the resulting determinant
using a regularization which respects
the invariance under Weyl transformations and
area-preserving diffeomorphisms.
(Obviously, in the diffeomorphism-invariant case a diffeomorphism-invariant
regularization of the determinant is instead appropriate.)}
\begin{eqnarray}
S^D= \frac{1}{2}\int\! dx\sss \left ( g^{\mu\nu}\partial_\mu\varphi
\partial_\nu\varphi - \alpha R(g)\varphi - 2 {\Lambda^D}\right )\ ,
\label{local2}\\
S^W = \frac{1}{2}\int \!dx \left( \gamma^{\mu\nu}\partial_\mu\varphi
\partial_\nu\varphi - \alpha R(\gamma)\varphi - 2 {\Lambda^W}\right)\ ,
\label{local1}
\end{eqnarray}
where $\varphi$ is an auxiliary scalar
field, and the parameter $\alpha$ is defined in terms of
$d$ by the relation $d = 1+ 12 \pi \alpha^2$.

The diffeomorphism-invariant action $S^D$ depends on four
independent fields\cite{abd91}: $\ggggg$, $\ggg$, $\gggg$, and $\varphi$.
However, in the Weyl-invariant case,
the fact
that ${\rm det} \gamma \! \equiv \! \aaaaa \aaa \! - \! \aaaa^2 \! = \! -1$
gives a relation between $\aaaaa$, $\aaa$, and $\aaaa$;
therefore there are only 3 independent fields, and we choose to
work with $\aaa$, $\aaaa$, and $\varphi$.

{}From Eq.(\ref{local2}), by first using integration by parts to obtain an
expression involving only first derivatives, and then
following a standard procedure of Legendre transform,
one finds the following first-order Lagrangian for the
diffeomorphism-invariant case
\begin{eqnarray}
{\cal L}^D= \int dx \Bigl(\pi^D_\varphi{\dot\varphi}
+\pi^D_{11}{\dot\ggg} - {\cal H}^D \Bigr) \, \label{lagr4}\\
{\cal H}^D =  - {\sss\over \ggg} {\cal E}^D + {\gggg\over\ggg} {\cal P}^D \ ,
\label{ham4}
\end{eqnarray}
where
\begin{eqnarray}
\pi^D_\varphi &=& {1\over \sss}(\gggg\varphi'-\ggg{\dot\varphi})
+{\alpha\over 2\sss}\Bigl({\dot \ggg}-2\gggg'+
{\gggg\over\ggg}\ggg'\Bigr) \, ,\label{pfd}\\
\pi^D_{11} &=& {\alpha\over 2\sss}\Bigl({\dot\varphi}-
{\gggg\over\ggg}\varphi'\Bigr) \, ,\label{paad}\\
{\cal E}^D &=&{1\over 2}\Bigl({\varphi'}^2
-{4\over \alpha^2}(\ggg\pi^D_{11})^2-
{4\over\alpha}\ggg\pi^D_{11}\pi^D_\varphi+ 2\alpha \varphi'' -
\alpha {\ggg'\over\ggg}\varphi' - 2 {\Lambda^D} \ggg
\Bigr)\ ,\\
\label{cons3}
{\cal P}^D&=&\pi^D_\varphi{\varphi'}-2{\pi^D_{11}}'\ggg-\pi^D_{11}\ggg'\ .
\label{cons4}
\end{eqnarray}
${\ggg}$ and $\varphi$ are dynamical fields, whose canonical momenta
are $\pi^D_\varphi$ and $\pi^D_{11}$,
whereas ${\sss/\ggg}$ and ${\gggg/\ggg}$ serve as
Lagrange multipliers for the
constraints\footnote{We use the
standard notation $A\! \sim \! 0$
to indicate that $A$ weakly vanishes, {\it i.e.}
$A$ vanishes on the constraint surface.}
${\cal E}^D \! \sim \! 0$
and ${\cal P}^D \! \sim \! 0$.
As it should be expected for this diffeomorphism-invariant theory,
both the Poisson brackets
of ${\cal E}^D$ and ${\cal P}^D$, which satisfy the
diffeomorphism algebra\cite{abd91},
and the Hamiltonian $\int \! dx {\cal H}^D$
vanish on the constraint surface.

The first order Lagrangian for the
Weyl-invariant case is given by
\begin{eqnarray}
{\cal L}^W= \int dx \Bigl(\pi^W_\varphi{\dot\varphi}
+\pi^W_{11}{\dot\aaa} - {\cal H}^W \Bigr) \, \label{lagr3}\\
{\cal H}^W =  - {1 \over \aaa} {\cal E}^W + {\aaaa\over\aaa} {\cal P}^W \ ,
\label{ham3}
\end{eqnarray}
where
\begin{eqnarray}
\pi^W_\varphi &=& (\aaaa\varphi'-\aaa{\dot\varphi})
+{\alpha\over 2}\Bigl({\dot \aaa}-2\aaaa'+
{\aaaa\over\aaa}\aaa'\Bigr) \, ,\label{pfw}\\
\pi^W_{11} &=& {\alpha\over 2}\Bigl({\dot\varphi}-
{\aaaa\over\aaa}\varphi'\Bigr) \, ,\label{paaw}\\
{\cal E}^W &=& {1\over 2}\Bigl({\varphi'}^2
-{4\over \alpha^2}(\aaa\pi^W_{11})^2-
{4\over\alpha}\aaa\pi^W_{11}\pi^W_\varphi+ 2\alpha \varphi'' -
\alpha {\aaa' \over \aaa} \varphi' - 2 {\Lambda^W} \aaa \Bigr)\ ,\\
\label{cons1}
{\cal P}^W&=&\pi^W_\varphi{\varphi'}-2{\pi^W_{11}}'\aaa-\pi^W_{11}\aaa'\ .
\label{cons2}
\end{eqnarray}
Here again there are two dynamical variables, ${\aaa}$ and $\varphi$,
whose canonical momenta are $\pi^W_\varphi$ and $\pi^W_{11}$; however,
there is only one lagrange multiplier, ${\aaaa/ \aaa}$, which
enforces the constraint ${\cal P}^W \! \sim \! 0$.

As expected based on the general arguments given at the beginning of
this section,
the Poisson brackets of ${\cal P}^W$ close on ${\cal P}^W$
\begin{eqnarray}
\{{\cal P}^W(x), {\cal P}^W(y)\}&=&[{\cal P}^W(x)
+ {\cal P}^W(y)]\delta' (x-y) \sim 0\ ,
\label{pwcloses}
\end{eqnarray}
and the Hamiltonian does not vanish
on the constraint surface defined by ${\cal P}^W \! \sim \! 0$;
in fact,
$\int \! dx {\cal H}^W \! \sim \! - \int \! dx ({\cal E}^W/\aaa)$.
However, the Dirac
Hamiltonian procedure for constrained systems\cite{dirac}
requires that the constraint surface be preserved by the time evolution,
{\it i.e.} the commutator of the Hamiltonian with the constraints
should weakly vanish, and instead one can show that
\begin{eqnarray}
\{{\cal H}^W(x), {\cal P}^W(y)\} \!&=&\!
\Bigl\{{1\!+\!\aaaa(x)\over \aaa(x)} , {\cal P}^W(y)\Bigr\}
{\cal P}^W(x)\! +\!
{1\!+\!\aaaa(x)\over \aaa(x)}
({\cal P}^W(x)\!+\!{\cal P}^W(y))\delta' (x\!-\!y)
\nonumber\\
& &\! + \Bigl[{1\over\aaa(x)}
({\cal P}^W(x)-{\cal E}^W(x))\Bigr]'\delta(x-y) \ ,
\label{weak2}
\end{eqnarray}
which reduces to $\{{\cal H}^W(x), {\cal P}^W(y)\}
\sim [-{\cal E}^W(x)/\aaa(x)]' \delta(x-y)$ on the constraint
surface defined by ${\cal P}^W \! \sim \! 0$.
It is therefore necessary\cite{dirac} to add a second constraint
$[-{\cal E}^W(x)/\aaa(x)]' \! \sim \! 0$, or equivalently
$-{{\cal E}^W(x)/\aaa(x)} \sim \Lambda^0$,
where $\Lambda^0$ is the constant mode of $-{\cal E}^W(x)/\aaa(x)$.
Enforcing this additional constraint with a Lagrange multiplier
$N(x)$, we obtain the Hamiltonian density
\begin{eqnarray}
{\cal H}^W =  - {1 \over \aaa} {\cal E}^W + {\aaaa\over\aaa} {\cal P}^W
- N ({{\cal E}^W\over\aaa} + \Lambda^0)
= -{N + 1\over \aaa}( {\cal E}^W + \Lambda^0 \aaa) +
{\aaaa \over \aaa}{\cal P}^W + \Lambda^0 \ .
\label{exten}
\end{eqnarray}

In this version of ${\cal H}^W$, $(N + 1)/\aaa$ serves as
Lagrange multiplier for the constraint ${\cal E}^W + \Lambda^0 \aaa \sim 0$,
and the Lagrange multiplier $\aaaa / \aaa$ enforces
again ${\cal P}^W\sim 0$.
Upon the identifications $\gamma_{11} \leftrightarrow g_{11}$,
$\gamma_{01} \leftrightarrow g_{01}$, $N+1 \leftrightarrow \sqrt{-g}$,
$\Lambda^D \leftrightarrow \Lambda^W - \Lambda^0/2$,
one can immediately see that the theory
with the Hamiltonian ${\cal H}^W$ in (\ref{exten})
is essentially equivalent to the theory
with the Hamiltonian ${\cal H}^D$ in (\ref{ham4}),
the only difference being the additional contribution to ${\cal H}^W$
from the constant mode $\Lambda^0$.
In particular, it is easy to verify that the constraints
${\cal E}^W + \Lambda^0 \aaa$ and ${\cal P}^W$
generate general ({\it i.e.} not necessarily area-preserving) diffeomorphism
transformations.

This result indicates that any quantization whose starting point is
the Dirac Hamiltonian
procedure would lead to equivalent theories for the two approaches.

\section{Anomalies and Ward Identities}
In Ref.\cite{pol87} the 1+1-dimensional gravity theory
defined by the diffeomorphism-invariant action (\ref{sgp}) is
investigated by using its anomalous Ward identities.  The starting
point is the functional integral
\begin{equation}
\label{func}
Z^D=\int {{\cal D} g \over \Omega_{diff}} ~
{\rm exp} (i \Gamma^D (g)) ~,
\end{equation}
where $\Omega_{diff}$ is the volume of the diffeomorphism group.

\noindent
In order to factorize out the gauge volume one may choose
to work in the light-cone gauge
$g_{--} \! = \!0$, $g_{+-} \! = \! 1$,
and introduce the corresponding action
for the ghost fields
\begin{equation}
\label{functionalgac}
Z^D \!= \!\! \int {\cal D} g_{++} {\cal D}c
{\cal D}b^{--} {\cal D}\chi_+ {\cal D}\chi_- ~
{\rm exp} \! \left [i \Gamma^D(g) + i \! \int \! d^2 \xi ~
(b^{--} \nabla_- \chi_- + c \nabla_+ \chi_- + c \nabla_- \chi_+)
\right ]
\end{equation}

After integrating out the ghost fields, $Z^D$ takes
the following form\cite{bil95}
\begin{equation}
\label{funct}
Z^D[J]=\int  {\cal D} g_{++} ~ {\rm exp}\left (i \Gamma^D(g)
+ i \Gamma^D_{gh}(g) \right ),
\end{equation}
where $\Gamma^D+\Gamma^D_{gh}$ is the gauge-fixed action for gravity.

In Ref.\cite{pol87}, in order to obtain the anomalous Ward identities,
the following infinitesimal shift in the functional variable of integration
is considered
\begin{equation}
\label{variation}
\delta_f g_{++}=(2 \partial_+ -g_{++}\partial_-) \, \delta f +
\delta f \, \partial_- g_{++} ~.
\end{equation}
The variation of the gauge-fixed action
$\Gamma^D+\Gamma^D_{gh}$  under this transformation can
be computed by exploiting the fact that the corresponding
energy momentum tensor
\begin{eqnarray}
\Theta^D_{\mu \nu}  \equiv {2 \over \sqrt{-g}}
{\delta [\Gamma^D(g) + \Gamma^D_{gh}(g)] \over \delta g^{\mu \nu}}
\label{tpgh}
\end{eqnarray}
satisfies the anomaly
relations\footnote{The form of these anomaly relations
can be derived from symmetry arguments\cite{bil95,kar94}, but
the value of the coefficient in front of $R(g)$ requires a calculation.
The value indicated in (\ref{anopgh}) follows from the fact that,
as shown
in Ref.\cite{pol87}, $\Gamma^D_{gh}(g) \! \sim \! -28 \Gamma^D(g)/d$.
Note that the $-28$ is the ghost contribution to the anomaly
in light-cone gauge, and is gotten\cite{pol87} by adding $-26$ for the
term $b^{--} \nabla_- \chi_-$
and $-2$ for the term
$c \nabla_+ \chi_- \! + \! c \nabla_- \chi_+$.}
\begin{eqnarray}
\nabla_\mu (g^{\mu \nu} \Theta^D_{\nu \alpha}) = 0  ~,
{}~~~~~g^{\mu \nu} \Theta^D_{\mu \nu} = 2 \Lambda^D + {d-28 \over 24 \pi} R(g)
{}~.
\label{anopgh}
\end{eqnarray}
{}From these relations
it follows that the variation of the gauge-fixed action
under the transformation (\ref{variation}) is given by
\begin{eqnarray}
\! \int \!\! d\xi^2
\frac{\delta [\Gamma^D(g) \! + \! \Gamma^D_{gh}(g)]}{\delta g_{++}}
\delta_f g_{++} \! = \!
\!\int \!\! d\xi^2
[ \nabla_\mu (g^{\mu \nu} \Theta^D_{\nu -})
\! - \! \frac{1}{2}\nabla_- (g^{\mu \nu} \Theta^D_{\mu \nu}) ]
\delta f
\! = \!\! \int \!\! d\xi^2
\frac{28 \! - \! d}{48 \pi}  \partial^3_- g_{++}  \delta f
\label{eqgac426}
\end{eqnarray}
which, following a standard
procedure\cite{bil95}, leads to the
anomalous Ward identities
\begin{equation}
\label{Ward}
\sum^n_i  \! \left < g_{++} \!  (\xi_1 ) \!
\dots \! \delta_{\!f} \! g_{++} \! (\xi_i ) \!
\dots  \! g_{++} \! (\xi_n)  \right > \!
+ \! \frac{d  \! \! - \! \!  28  \! \! +  \! \! \lambda^D}{i 48 \pi}
 \! \! \int \!  \! d\xi^2  \! \, \delta \! f \! ( \! \xi \! )
\left<  \partial_-^3 g_{++} \! (  \xi  )
g_{++} \! (\xi_1 ) \! \dots  \! g_{++} \!
(\xi_n)  \right > \!  \! = \!  \! 0.
\end{equation}
Here $\lambda^D$ is the additional contribution to the anomaly
which is due to the fact that $\delta_f g_{++}$
is a composition of a diffeomorphism
and a Weyl transformation on $g_{++}$,
and therefore the diffeomorphism-invariant
but not Weyl-invariant measure ${\cal D}g_{++}$ is not invariant
under $g_{++} \rightarrow g_{++} + \delta_f g_{++}$.
The direct evaluation of $\lambda^D$ is not known, but
a value of $\lambda^D$ can be fixed by requiring that
the theory be independent of the choice of gauge; specifically,
in Ref.\cite{pol87}
the class of gauges
$g_{--} \! = \!g_{--}^B$, $g_{+-} \! = \! 1$ is considered, and
it is found that the independence of the partition function on
the choice of $g_{--}^B$ requires
\begin{equation}
\label{lambda}
d-28+\lambda^D = {d-13- \sqrt{(d -1)(d - 25)} \over 2}~.
\end{equation}

We now turn to the Weyl-invariant
approach, and investigate its
anomalous Ward identities.
Let us start by considering the functional integral
\begin{equation}
\label{functional}
Z^W=\int {{\cal D} \gamma \over \Omega_{Sdiff}} ~
{\rm exp}(i \Gamma^W (\gamma)),
\end{equation}
where $\Omega_{Sdiff}$ is  the volume of the area-preserving
diffeomorphism group. The volume of the Weyl
group does not appear because the functional integral is already
written in terms of Weyl-invariant variables.
Fixing the $\gamma_{--} \! = \!0$ gauge, and introducing the corresponding
ghost action, one can rewrite $Z^W$ as
\begin{equation}
\label{functional2}
Z^W= \int  {\cal D} \gamma_{++} \,   {\cal D}c \,  {\cal D}b^{--} ~
{\rm exp} \left [i \Gamma^W(\gamma) + i \int \! d^2 \xi ~
b^{--}\hat\nabla_{-} (g_{- \alpha} \epsilon^{\alpha \beta}
\partial_{\beta} c) \right ] ~.
\end{equation}
Our choice of gauge is motivated by the fact that\cite{kar94,jac95}
in this gauge $\Gamma^W$ takes the same form as the light-cone-gauge
version of $\Gamma^D$, and we intend to exploit this correspondence
in the generalization of the Ward identities (\ref{Ward}) to the
Weyl-invariant approach.
Still, in the analysis we shall need to take into account
the fact that the form of the ghost action in (\ref{functional2})
is very different\footnote{One can understand
the pecular form of the ghost
action encountered in the Weyl-invariant approach
by noticing that the infinitesimal variation of
$\gamma_{\mu\nu}$ under an area-preserving diffeormorphism
can be locally written as
$\delta\gamma_{\mu\nu} \! = \!
\hat\nabla_\mu(g_{\nu \alpha}\epsilon^{\alpha \beta} \partial_\beta \phi)+
\hat\nabla_\nu(g_{\mu \alpha}\epsilon^{\alpha \beta} \partial_\beta \phi)$,
where $\phi$ is an arbitrary function.}
from the one in (\ref{functionalgac}),
and the measure ${\cal D}\gamma$, which is Weyl-invariant
but not diffeomorphism-invariant, is very different from
the measure ${\cal D}g$.

Integrating out the ghost fields in (\ref{functional2}), $Z^W$
takes the following form
\begin{equation}
\label{functional3}
Z^W = \int {\cal D} \gamma_{++} ~
{\rm exp} \left (i \Gamma^W({\gamma}) + i \Gamma^W_{gh}({\gamma}) \right ).
\end{equation}
The gauge-fixed action $\Gamma^W({\gamma})+\Gamma^W_{gh}({\gamma})$
that appears in (\ref{functional3})
has energy momentum tensor
\begin{eqnarray}
\Theta^W_{\mu \nu} ~ \equiv ~ {2 \over \sqrt{-g}}
{\delta [\Gamma^W(\gamma)+\Gamma^W_{gh}(\gamma)]
\over \delta g^{\mu \nu}} ~,
\label{tjgh}
\end{eqnarray}
which satisfies\footnote{Also the structure of these
anomaly relations is completely fixed by symmetry,
and we determined the coefficient $d \! - \! 28$ by observing
that
$\Gamma^W_{gh}(\gamma) \! \sim \! -28 \Gamma^W(\gamma)/d$
in $\gamma_{--} \! = \! 0$ gauge.}
the following anomaly relations
\begin{eqnarray}
\nabla_\mu (g^{\mu \nu} \Theta^W_{\nu \alpha}) = -
{d-28 \over 48 \pi} {1 \over \sqrt{-g}} \partial_\alpha R(\gamma) ~,
{}~~~~~g^{\mu \nu} \Theta^W_{\mu \nu} = 0 ~,
\label{anojgh}
\end{eqnarray}

In order to obtain the anomalous Ward identities,
in analogy with (\ref{variation}), we consider the following
shift in the functional variable of integration
\begin{equation}
\label{variationj}
\delta_f \gamma_{++}=(2 \partial_+ -\gamma_{++}\partial_-) \, \delta f +
\delta f \, \partial_- \gamma_{++} ~.
\end{equation}
{}From the anomaly relations (\ref{anojgh})
one can show that
the variation of the gauge-fixed action
$\Gamma^W+\Gamma^W_{gh}$  under the transformation (\ref{variationj})
is given by
\begin{eqnarray}
\! \int \!\! d\xi^2
\frac{\delta [\Gamma^W(\gamma) \!
+ \! \Gamma^W_{gh}(\gamma)]}{\delta \gamma_{++}}
\delta_f \gamma_{++} \! = \!
\! = \!\! \int \!\! d\xi^2
[ \nabla_\mu (g^{\mu \nu} \Theta^W_{\nu -})
\! - \! \frac{1}{2}\nabla_- (g^{\mu \nu} \Theta^W_{\mu \nu}) ]
\delta f \! = \!\! \int \!\! d\xi^2
\frac{28 \! - \! d}{48 \pi} \partial^3_- \gamma_{++}  \delta f .
\label{eqgac426bis}
\end{eqnarray}
The direct correspondence between (\ref{eqgac426bis}) and (\ref{eqgac426})
might be surprising considering that they were derived using very different
anomaly relations [(\ref{anojgh}) and (\ref{anopgh}) respectively];
however, we notice that in (\ref{eqgac426bis}) and (\ref{eqgac426})
the anomaly relations only appear in the
combination $\nabla_\mu (g^{\mu \nu} \Theta_{\nu -})
-\nabla_- (g^{\mu \nu} \Theta_{\mu \nu})/2$, and in the chosen gauges
this combination is essentially insensitive to
the difference between the
anomaly relations (\ref{anojgh}) and (\ref{anopgh}), since
\begin{eqnarray}
\nabla_\mu (g^{\mu \nu} \Theta^D_{\nu -}) = 0 ~, ~~~~
{}~~~~~~~~~~
{}~~~~~~
-\frac{1}{2}\nabla_- (g^{\mu \nu} \Theta^D_{\mu \nu}) =
\frac{28-d}{48 \pi} \, \partial^3_-  g_{++}  ~,
\label{bla1}\\
{}~\nabla_\mu (g^{\mu \nu} \Theta^W_{\nu -}) =
\frac{28-d}{48\pi} \, \partial^3_-  \gamma_{++}  ~, ~~~
-\frac{1}{2}\nabla_- (g^{\mu \nu} \Theta^W_{\mu \nu}) = 0 ~.
{}~~~~~~~
{}~~~~~~~~~~ \label{bla2}
\end{eqnarray}

Finally, from (\ref{eqgac426bis}) it is straightforward to derive
the following
anomalous Ward identities
\begin{equation}
\label{Wardj}
\sum^n_i  \! \left < \gamma_{++} \!  (\xi_1 ) \!
\dots \! \delta_{\!f} \! \gamma_{++} \! (\xi_i ) \!
\dots  \! \gamma_{++} \! (\xi_n)  \right > \!
+ \! \frac{d  \! \! - \! \!  28  \! \! +  \! \! \lambda^W}{i 48 \pi}
 \! \! \int \!  \! d\xi^2  \! \, \delta \! f \! ( \! \xi \! )
\left<  \partial_-^3 \gamma_{++} \! (  \xi  )
\gamma_{++} \! (\xi_1 ) \! \dots  \! \gamma_{++} \!
(\xi_n)  \right > \!  \! = \!  \! 0.
\end{equation}
The additional contribution $\lambda^W$ to the anomaly
is due to the fact that, as one can easily verify using
$\gamma^{\mu \nu} \equiv \sqrt{-g} g^{\mu \nu}$,
$\delta_f \gamma_{++}$ is an infinitesimal
(not area-preserving) diffeomorphism
transformation on $\gamma_{++}$,
and therefore the measure ${\cal D}\gamma_{++}$ is not invariant
under $\gamma_{++} \rightarrow \gamma_{++} + \delta_f \gamma_{++}$.
Also the direct evaluation of $\lambda^W$ is not known,
but we have verified that
the value of $\lambda^W$ obtained by imposing
the independence of this theory on the choice of gauge is equal
to the value of $\lambda^D$ analogously obtained for the conventional
diffeomorphism-invariant theory [see Eq.(\ref{lambda})].
This observation together with the results (\ref{Ward}) and (\ref{Wardj})
indicates that the anomalous Ward identities satisfied by $\gamma_{++}$
in the Weyl-invariant approach are identical to the ones
satisfied by $g_{++}$
in the diffeomorphism-invariant approach.
Since these Ward identities completely
determine\cite{pol87}
the Green's functions, also the Green's functions
are identical.

\section{Conclusion}
The results presented in the two preceding sections
indicate that, in spite of several
technical differences encountered along the way,
the two approaches are ultimately equivalent.

In the gauge-independent Hamiltonian formulation
it appears that the Weyl-invariant approach
leads to results that one would only
expect in the diffeomorphism-invariant approach.
In particular, by demanding the consistency of the
Dirac analysis of the Weyl-invariant case
one is led to the introduction of an additional independent
field, the Lagrange multiplier $N$,
which is related to the field $\sqrt{-g}$;
so it appears that
the action (\ref{sxg}) describes a $\sqrt{-g}$-dependent theory,
in spite of being $\sqrt{-g}$-independent ({\it i.e.}
involving $\gamma^{\mu \nu}$ only).

In deriving the equivalence of the two approaches at the level of
the anomalous Ward identities
a key role is played by the combination
$\nabla_\mu (g^{\mu \nu} \Theta_{\nu -})
-\nabla_- (g^{\mu \nu} \Theta_{\mu \nu})/2$,
which (in the chosen gauges) takes the same form in both approaches.
Clearly this combination encodes some essential feature of the theories,
but its physical interpretation is not yet clear to us.

We believe that (at least part of) the results
here found are a consequence of the
fact that the group of the area-preserving
diffeomorphisms is not invariant under arbitrary
coordinate redefinitions along the time direction;
for example, this leads to Eq.(\ref{weak2}).
It could be interesting to devise yet another quantization approach
with symmetry under transformations of a subgroup (of the diffeomorphism group)
that is invariant under coordinate redefinitions
along the time direction.

Further insight might be gained by analyzing the one-parameter family
of measures discussed in Ref.\cite{neu90}, which
interpolates between the two limiting cases considered here:
diffeomorphism invariance and Weyl invariance.
In the chiral Schwinger model a one-parameter ``$a$"-family
of chiral symmetry breaking measures
has also been identified\cite{jac85} and the mass emergent in that theory
depends on $a$; with two values of $a$ leading to the same mass.
It is conceivable
that the measures in the two approaches here considered are paired
in a similar fashion within the one-parameter family of measures
discussed in Ref.\cite{neu90}.

It would also be interesting to investigate some topological
issues which have been ignored here.
A differentiation
between the outcome of the two approaches
might be found if the topology of the space of metrics
$g_{\mu \nu}$ modulo diffeomorphisms
and the topology of the space of metrics
$\gamma_{\mu \nu}$ modulo area-preserving diffeomorphisms
do not coincide.

\bigskip
\bigskip
\bigskip
We would like to thank L. Griguolo and R. Jackiw for
very useful comments, and A. Alekseev and B. Zwiebach for
discussions.

\newpage
\baselineskip 12pt plus .5pt minus .5pt

\end{document}